# Title: Quantum Machine Learning with Application to Progressive Supranuclear Palsy Network Classification

Papri Saha[*]

Department of Computer Science, Derozio Memorial College, Kolkata-700136, India

ORCID: 0000-0001-9022-1947

**Abstract**: Machine learning and quantum computing are being progressively explored to shed light on possible computational approaches to deal with hitherto unsolvable problems. Classical methods for machine learning are ubiquitous in pattern recognition, with support vector machines (SVMs) being a prominent technique for network classification. However, there are limitations to the successful resolution of such classification instances when the input feature space becomes large, and the successive evaluation of so-called kernel functions becomes computationally exorbitant. The use of principal component analysis (PCA) substantially minimizes the dimensionality of feature space thereby enabling computational speed-ups of supervised learning: the creation of a classifier. Further, the application of quantum-based learning to the PCA reduced input feature space might offer an exponential speedup with fewer parameters. The present learning model is evaluated on a real clinical application: the diagnosis of Progressive Supranuclear Palsy (PSP) disorder. The results suggest that quantum machine learning has led to noticeable advancement and outperforms classical frameworks. The optimized variational quantum classifier classifies the PSP dataset with 86% accuracy as compared to conventional SVM. The other technique, a quantum kernel estimator, approximates the kernel function on the quantum machine and optimizes a classical SVM. In particular, we have demonstrated the successful application of the present model on both a quantum simulator and real chips of the IBM quantum platform.

*Corresponding author information: Papri Saha, Rajarhat Road, P.O.- R-Gopalpur, Kolkata-700136, India, (e-mail: saha.papri@gmail.com; pscompsc@dmc.ac.in).
2

# 1. Introduction

The advent of "resting-state" functional magnetic resonance imaging (rs-fMRI) has allowed researchers to visualize large-scale cortical networks in the human brain by mapping the regions-of-interest (ROI) in terms of temporarily correlated low-frequency fluctuations of fMRI signal that depends on the blood oxygen level [1, 2]. Many current studies have revealed that rs-fMRI-based network investigation is quite effective in identifying the unique signature of different neurological disorders [3-6]. Precisely, the functional brain network is a graph data structure with nodes representing the brain ROI and edges denoting the strength of the connection between those ROIs [7]. Therefore, the state-of-the-art diagnosis of any neurological disorder could emerge as a network classification problem. Imperatively, the primary task of such classification problems is the representation learning of graph-structured data. In particular, machine learning relates the feature space of factual data to discover generalized forms and intuitions with the aid of algorithms and statistical models without absolute instructions [8].

The intersection between machine learning (ML) and quantum computing has attracted significant interest in recent years [9-11] and several newly proposed quantum machine learning (QML) methods have been introduced to solve several real-life problems [12, 13]. The general approach assumes the initial problem of supervised learning: the creation of a classifier, where the network is presented with a labelled dataset $\chi = T \cup S = ((x_1, y_1), (x_2, y_2), ..., (x_m, y_m),) \subset R^d \times \{0,1, ..., c\}$. The training algorithm only considers the labels of the training data $T$. The objective is to formulate a map on the test set $S \to C$ where $C = \{0,1, ..., c\}$, such that it settles with high probability with the true map on the test set $s \in S$. Identification of a specific brain disorder from the fMRI datasets of patient and control groups is entirely based on a finite number of ROIs. For example, in the present work, where we have attempted to classify Progressive Supranuclear Palsy [14] (PSP)-affected networks from the control group, two reported classical pre-processing methods have used twenty-seven ROI datasets for the training vectors. One of the algorithms used the well-known eigenvector centrality (EVC) measure, while the other is based on a dimensionality reduction approach following Principal Component Analysis (PCA) [15]. Essentially, the objective of the second method was to find a new set of input data, smaller than the original one, nevertheless, retaining most of the original information. The primary target was to propose a map of binary classification based on the test set of PSP and control subjects. One such classical approach used the much-used support vector machines (SVMs) to construct an approximate binary labelling function [16]. Notably, SVM is a supervised algorithm that constructs a hyperplane $(w, b)$, parametrized by a normal vector $w \in R^d$ and $x_i \in T \subset R^d$ satisfying $w \circ x + b = 0$ associated to the inner product $\circ$ of vectors in $R^d$. The data set $\{x_i, y_i\}$ is linearly separable into a binary class $\{+1, -1\}$ by margin $2\|w\|^{-1}$ and subject to constraint $y_i(w \circ x_i + b) \geq 1$. Nonetheless, the real-world datasets (e.g., IRIS, PSP, breast cancer, etc.) are not linearly separable. Therefore, kernel function estimation is employed to transform the input dataset $\chi$ to a higher dimensional feature space that becomes increasingly expensive to compute in case of large feature data.

A quantum version of this method has previously been reported [17], where an exponential speed-up is implemented provided data is supplied in a coherent superposition. However, this is the Noisy Intermediate Scale Quantum (NISQ) technology age, where there exist only less than 100 qubits in a quantum processor [18]. Thus, hybrid quantum-classical architecture approaches are often used in the existing QML algorithms. One of the approaches uses a variational quantum circuit [19] that is designed to create a separating hyperplane in the respective quantum state space followed by a binary measurement. Another approach builds the hyperplane using a classical SVM, only using the quantum machine to evaluate the kernel function. Moreover, to improve over the classical approaches, a map was designed based on a parameterized quantum circuit that is quite difficult to simulate classically but can be realized even on noisy quantum devices. The input data was non-linearly mapped to a quantum feature state $\Phi: x \rightarrow |\Phi(x)\rangle\langle\Phi(x)|$. The state fidelity was used to estimate the similitude between the output label and the actual label of quantum states [20]. Additionally, the parameters of the variational circuit were periodically updated as well. Imperatively, the process of modifying the hyper-parameters to train the parameterized quantum circuit in assigning the correct label is nothing but optimization. The set of optimal hyper-parameters guarantees better resolutions in classification problems.

The objective of the present study is to propose an optimized quantum classifier that must result in superior outcomes relative to the existing ML algorithm. However, all network datasets were subject to PCA a priori for the sake of dimensionality reduction. The outcomes are apprehended to upgrade the state-of-the-art PSP diagnosis as the proposed QML was noted to offer a better approach to leverage the complex brain network data in drawing conclusive inferences about the unique connectivity signature of PSP subject groups and also to improve brain network classification techniques applicable for a wide variety of neuropathological conditions. The key contributions of the present work are outlined below:

(i) We have proposed a graph-supervised learning for brain network analysis capable of better feature recognition from a limited data resource and offered a superior diagnostic tool. To the best of our knowledge, the present work is a first-of-its-kind attempt at employing QML for brain network analysis from a graph-theoretic perspective. The outcomes holistically illustrate the importance of QML in brain network classification problems.

(ii) We have used the state-of-the-art quantum computer made available by the IBM quantum experience platform [21] in both simulator and real chip mode to implement our algorithm. We have considered the PSP dataset consisting of eight features and a four-qubit encoding procedure was created thereof.

The article is structured as follows. In Sec. 2, we have presented a brief methodology to implement EVC, PCA, variational quantum circuit, and quantum SVM (QSVM) algorithm. Sec. 3 is dedicated to the application of QML as well as to classical pre-processing methods. The proposed method was observed to outperform the classical SVM algorithm. We have implemented the QML algorithm on both the simulator and the real chip. We have also applied the error mitigations to ensure reliable calculations. A summary and conclusion are presented in Sec. 4 and 5.

## 2. Methods

2.1 Brain network data and the chosen network measure for dimensionality reduction

The brain functional connectivity dataset, used in the present study, was obtained from the online repository known as "USC Multimodal Connectivity Database (UMCD)" (http://umcd.humanconnectomeproject.org/). The brain connectivity data, obtained by processing the fMRI signals at different levels, is typically presented in the form of a graph data structure specifying interconnections among different ROIs. According to the methodology outlined in UMCD, the PSP brain network data, localized over the rostral midbrain tegmentum region, was derived for 27 nodes by applying a regional mask to the raw fMRI signals and finally stored in the form of a $27 \times 27$ connectivity matrix $A$ for each subject. The details of each ROI are given under supplementary Table S1. Each element of matrix $A$ denoted the pairwise correlation strength between two ROIs [22]. Formally, the $n \times n$ correlation matrix $A$ where $n$ denotes the number of nodes, characterizes the distribution of edge weight $\{a_{ij}\}_{i=1,n;j=1,n}$ between all node pairs. However, according to the present PSP data structure, any protocol based on such a connectivity matrix must be a three-dimensional analysis as the two-dimensional connectivity matrix is to be combined with the one-dimensional vector of members present in different subject groups. We have worked with male and female subject groups and each group comprised of healthy control (HC) and PSP-affected members. The PSP subject's dataset was equidistributed in female and male groups with 12 members in each group. The HC dataset consisted of 16 males and 24 females. A summary of the dataset is given in Table 1.

**Table 1:** Summary of the brain connectivity dataset

|  | Male | | Female | |
| --- | --- | --- | --- | --- |
|  | Heathy Control (HC) | PSP-affected | Heathy Control (HC) | PSP-affected |
| Count | 16 | 12 | 24 | 12 |
| Age (in years) | 69.32±13.58 | 69.16±12.6 | 67.76±4.44 | 67.26±5.84 |

Initially, we have reduced the dimensionality of the input dataset to optimize the computational overheads. One of the dimensionality reduction approaches may be to convert the two-dimensional connectivity data of individual members to a one-dimensional node-wise vector of an appropriate network measure. Eigenvector centrality (EVC), a network measure signifying the relative importance of a node in overall connectivity structure, may be used for such dimensionality reduction [23]. Imperatively, EVC of brain ROIs have been proved to be useful in studying various neuropathological disorders including Alzheimer's [3], Parkinson's [24], ASD [25], intractable focal epilepsy [26], and longstanding type-1 diabetes [27]. EVC, unlike other centrality measures (e.g., degree, closeness, betweenness), considers the connectivity information of adjacent nodes together with the node of interest in assigning the centrality score, which signifies if a node is connected to a set of highly connected nodes, it will have a high EVC score. The formal definition of EVC is described as follows:

Let the centrality value of a node $k(a_k)$ is given as

$$a_k = \frac{1}{\lambda}\sum_{i \in M(R)} A_{k,i} a_i, \qquad (1)$$

where $M(R)$ is the set of adjacent nodes of $k$. The equation could further be simplified to a matrix notation as

$$Ax = \lambda x \qquad (2)$$

Thus, the centrality value may be characterized in terms of eigenvector $x$ of $A$. For positive-valued eigenvector (i.e., $x_k > 0, \forall\ k = 1, n$), the eigenvector accompanying the maximum eigenvalue (i.e., $\lambda = \lambda_{max}$) was defined as the EVC according to the Perron-Frobenius hypothesis [28, 29].

Applying the node-wise EVC measure, the connectivity matrix may be transformed to a vector and the dimensionality of the input dataset was reduced to two. Thus, the male group's original dataset of size 28 (number of subjects) × (27 × 27) (size of the connectivity matrix) was reduced to 28 (number of subjects) × 27 (size of the EVC vector). Further, to avoid the higher number of features and not compromise the performance, we have carried out the dimensionality reduction by Principal Component Analysis.

### 2.2 Principal Component Analysis

Principal Component Analysis (PCA) uses fundamental mathematical techniques that transform a number of potentially correlated variables into a smaller number of uncorrelated variables referred to as "principal components". The primary task of PCA is to detect a completely new set of orthogonal coordinates from the original data. This was accomplished by investigating the orientation of maximal variance along the coordinates of 28 × 27 and 36 × 27 two-dimensional EVC-member datasets of male and female groups, respectively. Mathematically, for any input dataset $X$ with $n$ observations and $d$ features, PCA first generates another matrix $Y$ of dimension $n \times d$ through a linear transformation operator $P$ of dimension $d \times d$ as

$$Y = PX \tag{3}$$

where, rows of $P$ (i.e., $p_1, p_2, ..., p_d$) become the new basis set for defining the columns of $X$ (i.e., $x_1, x_2, ... x_n$) and $(p_i \circ x_j), p_i, x_j \in R^n$ is the standard Euclidean inner product. Another diagonal matrix $S_Y \left(= \frac{1}{n-1} YY^T\right)$ is also introduced, which represents the covariance of the mean-centred data. Precisely, the covariances between separate measurements are reduced to zero by diagonalizing $S_Y$ to remove any redundancy present in the input data. Moreover, PCA assumes $P$ is an orthonormal matrix and the directions with the largest variances are the most "important" or in other words, the principal ones. Rewriting $S_Y$ in terms of the chosen variable $P$ and using equation 3, we have

$$S_Y = \frac{1}{n-1} YY^T = \frac{1}{n-1}(PX)(PX)^T = \frac{1}{n-1} PXX^T P^T = \frac{1}{n-1} PAP^T \tag{4}$$

Here, we defined a new matrix $A = XX^T$ which is symmetric, and is diagonalized by an orthogonal matrix of its eigenvector $E$, such that $A = EDE^T$. Upon considering $P = E^T$, we calculate $A = P^T DP$. Therefore, using the property that the inverse of an orthogonal matrix is its transpose, we finally have

$$S_Y = \frac{1}{n-1} PAP^T = \frac{1}{n-1} P(P^T DP)P^T = \frac{1}{n-1} D \tag{5}$$

First $k < d$ principal components with maximum variances correspond to the reduced feature set. We have applied several classical as well as quantum ML algorithms on these feature variables to differentiate PSP from HC subject groups.

### 2.3 PSP classification using classical ML algorithm

Our primary objective of this study was to design and implement an efficient quantum ML method to classify PSP subjects. Nevertheless, we have also tested the efficacies of some popular classical ML algorithms including decision trees, linear regression, logistic regression, k-nearest neighbors (KNN), support vector machines (SVM), random forest, naïve Bayes, and neural networks [6, 8] for comparison purposes. All ML algorithms were applied upon splitting the feature set into two groups, namely (i) the training (size: 70%) and (ii) validation (size: 30%). The step-wise algorithm to implement the classical SVM algorithm starting from the baseline correlation matrices is as follows:

---

**Algorithm 1**: *PSP* classification algorithm

**Input**: *N*: Regions of Interest, *A*: $N \times N$ correlation matrix, $(x, y, z)$: *MNI* coordinates

*for s* = 1 *to* 2 *do*          // *s* can be male or female

    *for k* = 1 *to* 2 *do*          // for *PSP* and *HC* group

        *for i* = 1 *to* $N_{x,y,z}$ *do*

            *for j* = 1 *to m do*   // *m* is the total number of subjects in a specific category

                calculate $EVC_{i,j}$ // $EVC_{i,j}$ is the eigenvector centrality measure

            *end*

        *end*

    *end*

    *train*_features ← *dataset* (*EVC*, *prob* = 0.70)

    *test*_features ← *dataset* (*EVC*, *prob* = 0.30)

                // splits the entire dataset into training and test feature and label sets

    *pca_train* = *pca.transform* (*train_features, n_dim*)

    *pca_test* = *pca.transform* (*test_features, n_dim*)

                // evaluate *n*-dimensionality reduced *PCA* components over *N*

    *svc* = *SVC* (*pca_train, train_labels*)

    *train_score* = *svc.score (pca_train, train_labels)*

    *test_score* = *svc.score (pca_test, test_labels)*

                // do classical *SVM* on the training as well as on the test dataset

*end*

**Output**: *svc*

In the following, we define two quantum classifiers to differentiate PSP subjects from the healthy control (HC) group. The first classifier is based on a variational circuit [19, 30] that creates a differentiating hyperplane in the quantum state space. The second one employs the quantum computing machine to calculate the kernel function of the quantum feature space and then simulate a classical SVM.

### 2.4 Quantum Variational Circuit

In the QML paradigm, a quantum feature map transforms $x \in R^d$ into a quantum state $|\Phi\rangle\langle\Phi|$ on an $n$-qubit registrar. Here $H = C^2$ is a single qubit complex vector (Hilbert) space, and $Q(H^{\otimes n})$ signifies the cone of positive semidefinite density matrices $\rho \geq 0$ with unit trace $tr[\rho] = 1$. The transformation takes place using a unitary matrix form $U_\Phi(x)$, which is typically called a "parameterized quantum circuit" or "quantum variational circuit" (QVC) [19]. Their parameters are decided by the data being encoded and the optimization process. To acquire quantum supremacy, we could do with these maps to result in a kernel $K(x, y) = |\langle\Phi(x)|\Phi(y)\rangle|^2$ that is mathematically hard to evaluate in a classical paradigm. Again, they can generate a key subspace of states in the output Hilbert space that allows them to be used as a machine learning classifier model. The quantum feature map with depth $d$ can be defined as follows

$$U_\Phi(x) = \prod_d U_\Phi(x) H^{\otimes n}, \qquad (6)$$

$$U_\Phi(x) = exp\left(i \sum_{k \subseteq [n]} \Phi_k(x) \prod_{m \in k} P_i\right), \qquad (7)$$

Essentially, the map contains a series of Hadamard gates interwoven with entangling blocks $U_\Phi(x)$. Within the entangling blocks, $U_\Phi(x): P_i \in \{I, X, Y, Z\}$ represents the Pauli matrices and the index $k$ describes the connectivity among different qubits or data points. The $2^n$ possible coefficients of $\Phi_k(x) \in R$ are the non-linear mappings of the input data $x \in R^d$. Specifically, in the present manuscript we have considered $d = 2, P_0 = Z, P_1 = ZZ$, which is the second-order

Pauli-Z evolution circuit termed as *ZZFeatureMap* circuit in Qiskit [31]. For both the training and the testing (classification) phases, the algorithm executing QVC consists of three major sections: the encoding of the quantum feature space, the variational optimization, and the output measurement. These steps are then combined into the circuit representation depicted in Fig. 1 (b).

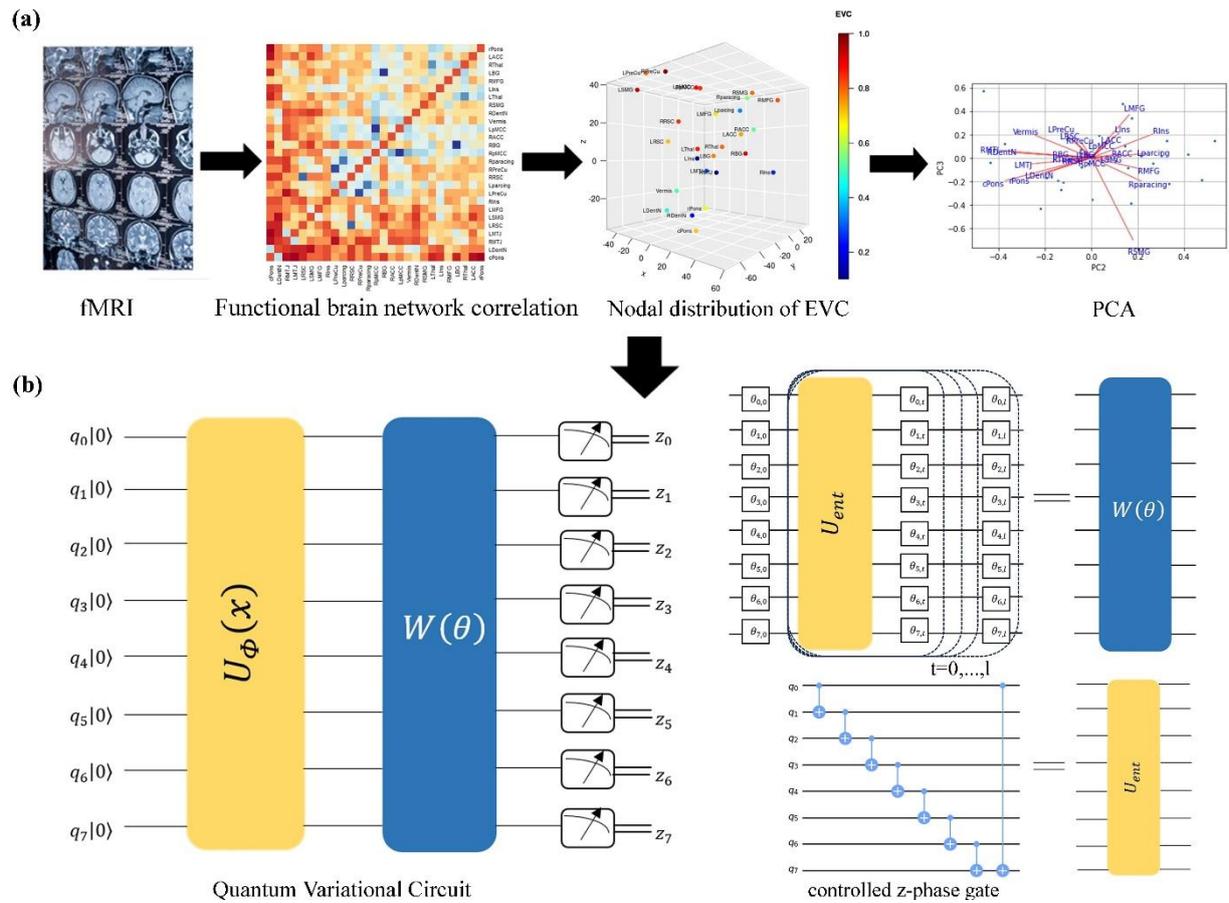

**Fig. 1.** Overview of the methodology. **(a)** Data preprocessing steps: an fMRI image taken from https://doi.org/10.1002/mds.28087; correlation matrix representing a PSP brain network of http://umcd.humanconnectomeproject.org/ ; nodal distribution of EVC in a three-dimensional scatter plot format; 2-component PCA representation for 27 ROIs; **(b)** Quantum variational optimization method with commonly used ansatz for the variational unitary (repeated $l$ times) as $W(\theta) = U_{loc}^{(l)}(\theta_l) U_{ent} \ldots U_{loc}^{(2)}(\theta_2) U_{ent} U_{loc}^{(1)}(\theta_0)$. The entangling gates $U_{ent}$ consisted of controlled-Z phase gates $CZ(i,j)$ with full layers of single-qubit rotations $U_{loc}^{(t)}(\theta_t) = \otimes_{i=0}^{n-1} U(\theta_{i,t})$.

The QVC was designed to receive the reference state, $|0\rangle^n$ as input. Subsequently, it employs the unitary $U_\Phi(x)$ and the variational unitary $W(\theta)$ operations and finally, generates the output measurement in canonical Z-basis. The output bit string $z \in \{0,1\}^n$ is then mapped to a label in $C$. This circuit is then repeated multiple times and sampled to finally approximate the expectation values of each Pauli term in the Hamiltonian. Each measurement $\{M_y\}_{y \in C}$ corresponds to the output labels $y$. In the present work, we have considered $C = \{0,1\}$ that correspond HC and PSP, respectively. The QVC based PSP classification scheme is elaborated as follows:

---

**Algorithm 2**: *PSP* classification using *QVC* model

**Input**: *n_dim, pca_train, train_labels, pca_test, test_labels* // training and test feature and label sets

*for s* = 1 *to* 2 *do*          // *s* can be male or female

    *feature_map = ZZFeatureMap* (*feature_dimension=n_dim*)

                      // create *ZZFeatureMap* circuit using standard *Qiskit* library

    *ansatz = RealAmplitudes*(*num_qubits=n_dim*)

                      // create a variational quantum *ansatz* of *n* dimensional qubits

    *vqc = VQC* (*feature_map, ansatz, optimizer=COBYLA*)

                      // create *VQC* using *COBYLA* optimizer

    *vqc.fit* (*pca_train, train_labels*)

    *train_score = vqc.score* (*pca_train, train_labels*)

    *test_score = vqc.score* (*pca_test, test_labels*)

                      // do *QVC* on the training as well as on the test dataset

    *metrics* ← *classification_report* (*test_labels, vqc.predict* (*pca_test*))

    // *metrics* evaluated using confusion matrices with *accuracy, sensitivity, specificity*, etc. parameters

*end*

**Output**: *vqc, metrics*

## 2.5 Quantum Kernel and Quantum Support Vector Machine

A quantum support vector machine (QSVM) may be considered a classical SVM with a quantum kernel. A common approach to defining a quantum kernel is to consider the fidelity of two different inputs of the feature map circuits that correspond to the Hilbert-Schmidt inner product in the feature space. Considering that the feature map is a parameterized quantum circuit we can calculate kernel matrix with $n$-qubits for every pair of data points in the training dataset $x_i, x_j$ as [32, 33]

$$K_{ij} = K(x_i, x_j) = |\langle \Phi(x_j)|\Phi(x_i)\rangle|^2 = \left|\langle 0^{\otimes n}|U^\dagger_{\Phi(x_j)} U_{\Phi(x_i)}|0^{\otimes n}\rangle\right|^2 \qquad (8)$$

It follows that the output probability could only be estimated by sampling the expectation of output measurements with $R$ shots and considering only the $0^n$ count. Once the kernel matrix with complete training data is developed, the conventional SVM classifier is chosen to operate. A linear SVM is designed to find a hyperplane that distinguishes the data, with the maximum attainable distance between the two sets. The longitudinal distance between the hyperplane and two data points with varied labels is termed the 'margin' that needs to be maximized subject to a constraint $y_i(w \circ x_i + b) \geq 1$ mentioned earlier and such points are referred to as the "support vectors". The respective cost function is given as [33]:

$$L_P = \frac{1}{2}\|w\|^2 - \sum_{i=1}^{t} \alpha_i y_i(w \circ x_i + b) + \sum_{i=1}^{t} \alpha_i \qquad (9)$$

where $\alpha_i \geq 0$ are the Lagrange multipliers.

It is useful to formulate the dual of the original primal problem $L_P$ presented in eqn. (9), when a much higher dimensional feature space with $d \ll dim(H)$ for the vectors $\Phi(x)$ with $x \in T$ is considered and is expressed as follows [33]:

$$L_D = \sum_i \alpha_i - \frac{1}{2}\sum_{i,j} \alpha_i \alpha_j y_i y_j \, x_i \circ x_j, \; 0 \leq \alpha \leq C, \sum_i \alpha_i y_i = 0 \qquad (10)$$

It is followed from eqn. (8) that an optimal hyperplane is calculated following the dual problem $L_D$ in eqn. (10) that is fully defined after we have been given the output labels $y_i$ and have approximated the kernel $K(x_i, x_j)$.

In our classification stage, we want to determine a label for a new datum $s \in S$ of the test set. Therefore, the inner product $K(x_i, s)$ among all support vectors with $x_i \in T$ and the new datum $s$ are to be evaluated on a quantum computer. Specifically, if we can evaluate the kernel $K(x, y) = \Phi(x) \circ \Phi(y)$ [32, 33], we can construct a classification function as

$$m(s) = sign\left(\sum_{i \in N_S} \alpha_i y_i K(x_i, s) + b\right) \qquad (11)$$

where the summation is considered over all support vectors $i \in N_S$ such that $\alpha_i \geq 0$ and the label can be directly computed from the kernel $K(x_i, s)$.

Depending on the dataset, the QSVM is robust against noise. In general, an error mitigation technique is applied to ensure reliable computations.

All computations were performed in Qiskit-an open-source software developed by IBM (IBM Quantum Experience Platform).

## 3  Results

### 3.1 Data preprocessing

Initially, the functional connectivity information was extracted in terms of their correlation matrices (i.e., *A*). Subsequently, the nodal distribution of the EVC measure was used to represent the data for 27 ROIs [34]. However, when PCA is applied, it is often necessary to analyse how much EVC data variation is contained by each principal component. This is fundamental because it acknowledges ranking the components in order of significance and centres on the highly important ones when analysing the results of our study with fewer parameters. Accordingly, the explained variance ($S_Y$) will then be considered in choosing the number of dimensions to keep in an EVC-reduced dataset. Fig. 2 shows the PCA analysis for the PSP-female dataset. A similar analysis for PSP-male is given in the supplementary section. The blue bars depict the percentage variance described by each principal component (follows *PCA.explained_variance_ratio_*), whereas the red line displays the cumulative sum (follows *PCA.explained_variance_ratio_.cumsum()*). Fig. 2 a indicates the percentage of the explained variance in the given data as the principal components are gradually added. The first principal component (PC1) describes 32% of the variance of the PSP data set. Similarly, the first two principal components describe 61%, the first three explain 76%, the first four explain 89%, and so on. The figure shows that it is enough to consider only 8 out of the 27 principal components to describe 96% of the explained variance in the EVC-reduced PSP dataset. Fig. 2 b displays a heatmap showing connectivity correlation among 8 principal components distributed over 27 brain regions. It follows from the diagram that once the PCA was applied, one can also have the components as principal axes in output feature space, representing the associations of maximum variance in the dataset. We can plot these component scores on an initial variable factor map chart (Fig. 2 c) to impact the influence of input feature sets on the final principal components. Each plot is drawn on a factorial plane such that the vector space consists of the intersection of two of the principal components. Firstly, we plot the component scores of PC1 against PC2: the distance of the line segments signifies how much the brain ROI describes the variance of the data on the factorial plot. The angle within the different ROIs manifests how well the regions were correlated. The details for all factorial planes are depicted in Fig. 2 c. The supplementary figure S1 depicts the same for the PSP-male dataset. Thus, the principal components derived from this unsupervised learning approach take values as dimensionality-reduced components of the 27 brain ROIs.

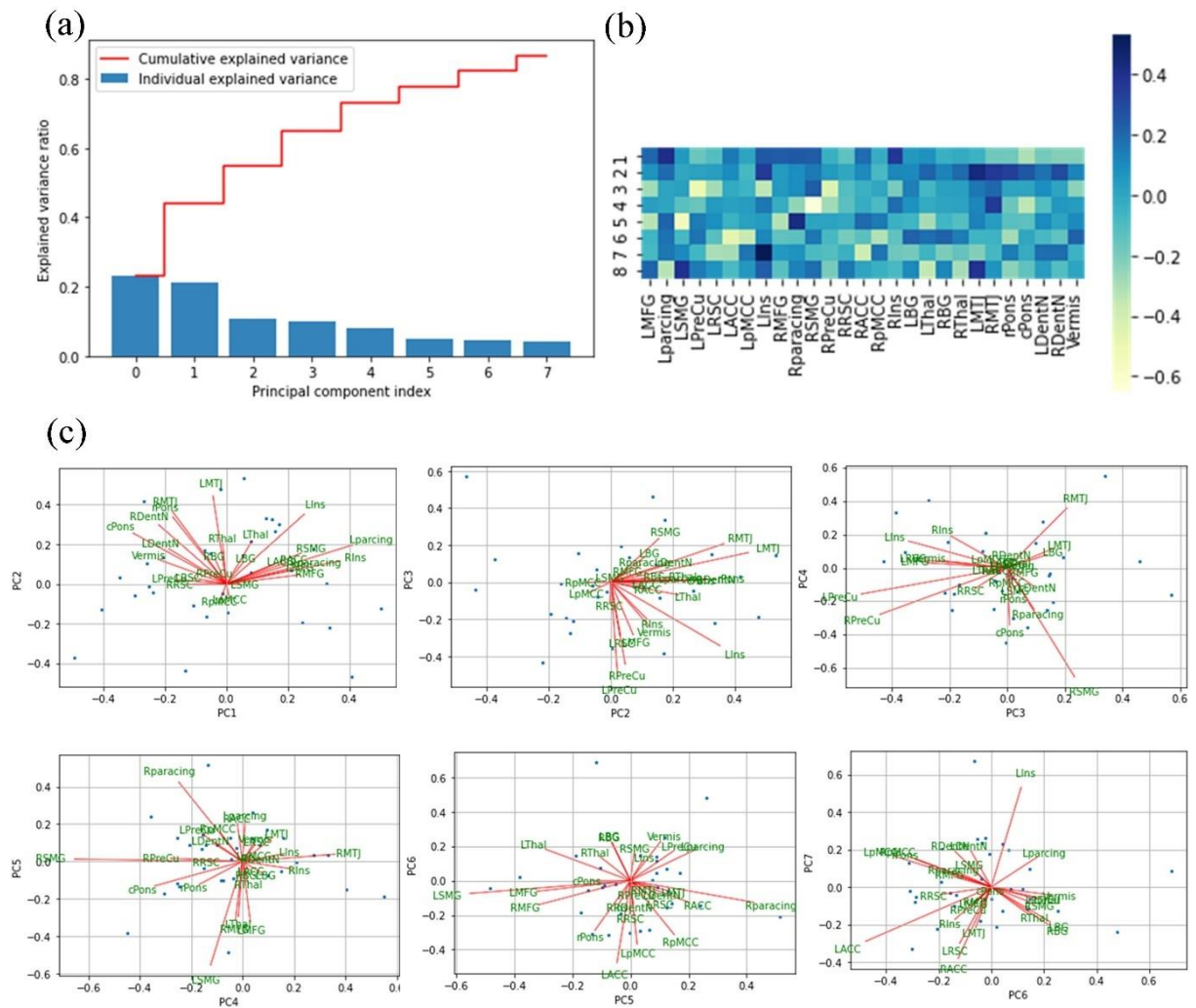

**Fig. 2.** PCA on the EVC reduced PSP female dataset. **(a)** Scree plot for proportion of variance explained, **(b)** Correlation heatmap among 8 principal components distributed over 27 brain ROIs, **(c)** Variances of ROIs on the factorial plane. Each plane consisting of the intersection of two of the principal components.

In the following, we describe two binary classifiers that will classify HC from the PSP subject group based on EVC and PCA-reduced dataset considering a quantum state space as feature vector space. Here, the input data was transformed non-linearly into a quantum state space $\Phi: x \in R^d \to |\Phi(x)\rangle\langle\Phi(x)|$. The first approach, the quantum variational classifier, employs a variational quantum circuit [31, 35] to classify the PSP data, which is analogous to the conventional SVMs and creates a linear decision mapping in quantum feature space. The second procedure, a quantum kernel estimator, evaluates the kernel function on the quantum machine and optimizes a classical SVM. Results showed the consequences of these two methods of QML in PSP network dataset classification.

### 3.2 QVC on PSP dataset

Training and classification with a typical SVM are considered to be successful when the inner products between feature vectors could be calculated effectively [36, 37]. Classifiers relying on quantum circuits (QVC), for instance, the one given in Fig. 1 b, cannot deliver a quantum privilege over a conventional SVM if it is possible to evaluate the feature kernel $K(x,z) = |\langle\Phi(x)|\Phi(z)\rangle|^2$ on a classical machine. To extract the advantage from classical methods, we have chosen a function that is based on short-depth circuits (Eqs. 6-7) and also compliant with error-mitigation techniques subject to the event of minimal decoherence. Imperatively, the QVC was designed to operate in succession of four steps (Algorithm 2). Primarily, the input data $x \in R^d$ was transformed to an 8-qubit quantum state by applying the *ZZFeatureMap* circuit $U_\Phi(x)$, where $\Phi$ is a non-linear function and $\Phi(x,y) = (\pi - x)(\pi - y)$, which defaults to $\Phi(x) = x$ as in Fig. 3 a. Next, a short-depth circuit (ansatz) $W(\theta)$ was chosen in the respective quantum feature space. The circuit with two layers was parameterized by $\theta$ in $Y$ rotations and controlled-NOT (CNOT) gates and was subject to optimization during the training phase. The CNOT gates confirmed the entanglement of formation amongst the encoded inputs. Two entangling layers (Fig. 3 b) were given to certify that the states of all input qubits become associated with each other and that quantum computational gain could be accomplished. In the third step, a binary measurement $\{M_y\}_{y \in \{+1,-1\}}$ diagonal in the Pauli z-basis (Fig. 1 b), was applied to the state $W(\theta)U_\Phi(x)|0\rangle^n$. Finally, in the fourth step, we examined $R$ repetitive measurement shots using Constrained Optimization By Linear Approximation optimizer (COBYLA) to obtain the empirical distribution as in Algorithm 2 (Fig. 3 c). The training and testing datasets comprised 22 and 10 points, respectively. These were obtained after removing "outliers" in the initial dataset by calling *scipy.stats.zscore(df)* and following *(array < 3).all(axis=1)* with *df* as a *DataFrame* (*NumPy* array) containing the z-score of each value in the sample, relative to the mean and standard deviation. The classification accuracy was found to be 86% in this environment. When the circuit depth is 2, much lesser resources are required to develop the circuit. Nonetheless, it is assumed to get better classification success for increased depth. For any binary classification task, the accuracy could be calculated in terms of positives and negatives as: $accuracy = \frac{TP+TN}{TP+TN+FP+FN}$, where $TP =$ True Positives, $TN =$ True Negatives, $FP =$ False Positives, and $FN =$ False Negatives. Furthermore, to better analyse the performance of the QVC model, a closer investigation of true positive $\left(\frac{TP}{TP+FN}\right)$ and false positive rates $\left(\frac{FP}{FP+TN}\right)$ in terms of a typical Receiver Operating Characteristic (ROC) curve at all classification thresholds was plotted. Furthermore, AUC (i.e., Area Under the ROC Curve), which supposedly indicates a cumulative effect over all probable classification thresholds, was found to be as high as 0.85 in our case (Fig. 3 c insert). However, AUC alone could not provide the whole picture, especially in cases where a substantial disparity between the positive and negative label counts exists. Henceforth, we also look at two other metrics called precision and recall. A complete classification report for the QVC classifier is presented in Fig. 3 d.

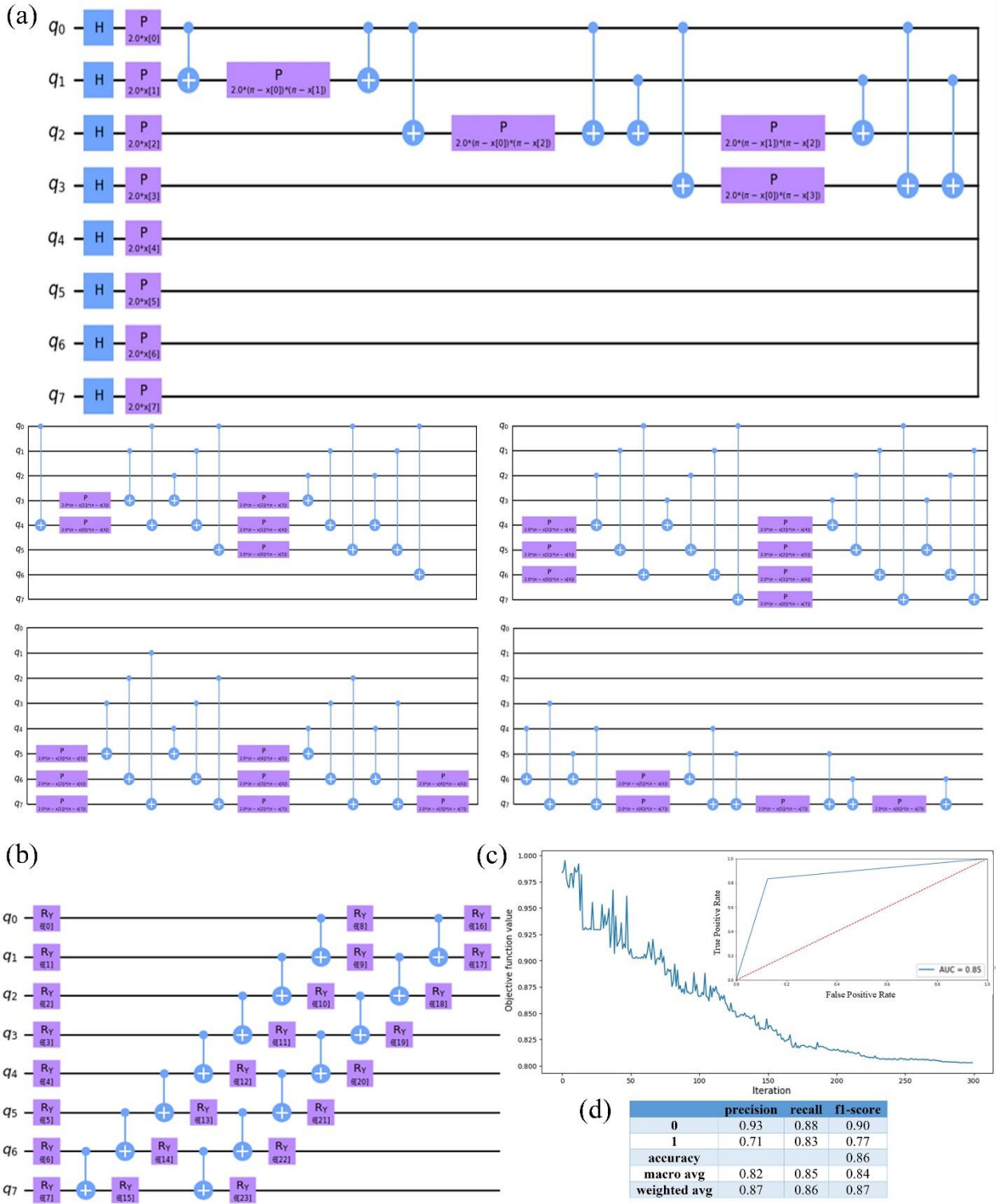

**Fig. 3.** QVC on PCA reduced PSP-female dataset. **(a)** 8-qubit *ZZFeatureMap* circuit with depth 1, **(b)** 2 layers ansatz, **(c)** COBYLA optimizer with $R = 300$ shots (insert ROC graph for variational quantum circuit), **(d)** classification report.

Another important performance analysis issue of a QML algorithm is the upgradation achieved relative to classical machine learning or similar classical algorithms. Therefore, the present work also includes the testing of different classical ML algorithms together with SVM to envisage the probable computational benefit in the application of QVC to the PSP dataset classification (Fig. 4).

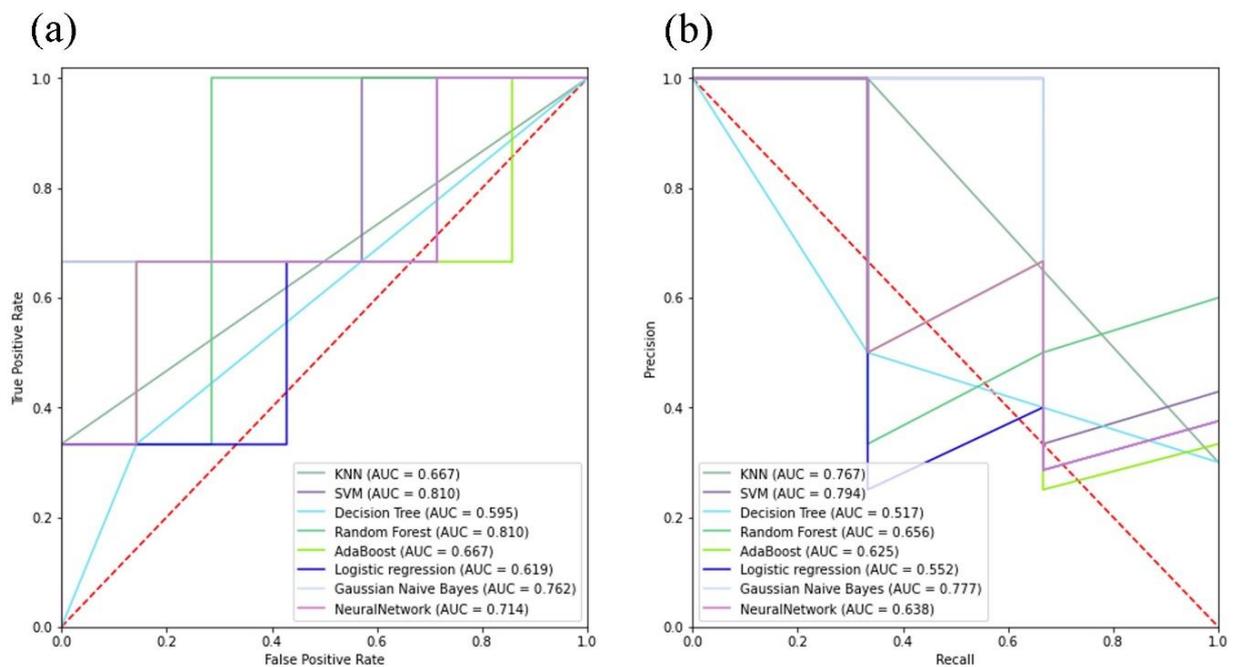

**Fig. 4.** Classical ML algorithms after PCA reduced PSP-female dataset. **(a)** Receiver operating characteristic (ROC), **(b)** Precision-Recall curve.

### 3.3 Quantum kernel function

The next classification algorithm, namely, the quantum kernel estimation, used the same training data to execute the kernel SVM. Likewise, the dataset was created from the *ZZFeatureMap* with SVM's $depth = 2$ and $dimension = 8$. The dataset consisted of 2 classes, with 22 training and 10 testing data points from each class. Initially, we have calculated the training and testing quantum kernel matrices following equation (8) as shown in Fig. 5. We have used an instance of *FidelityQuantumKernel* that holds the feature map along with its parameters to train our quantum kernel. We then passed the trained kernel to a machine-learning model. Subsequently, the *fit* method was invoked to test the corresponding performance for new data. Here, we have used the well-known Qiskit Machine Learning's *SVC* for classification [31]. This classification results in a success of 76% only. The final kernel matrix displayed the measure of similarity between the training samples. We have observed that the precision acquired on a quantum kernel estimation was nearly approaching that attained on a classical ML algorithm that is anticipated to produce slight time advancement over the classical computer (Fig. 5 c). In particular, there exists a trade-off of accuracy and shots with these training data sets.

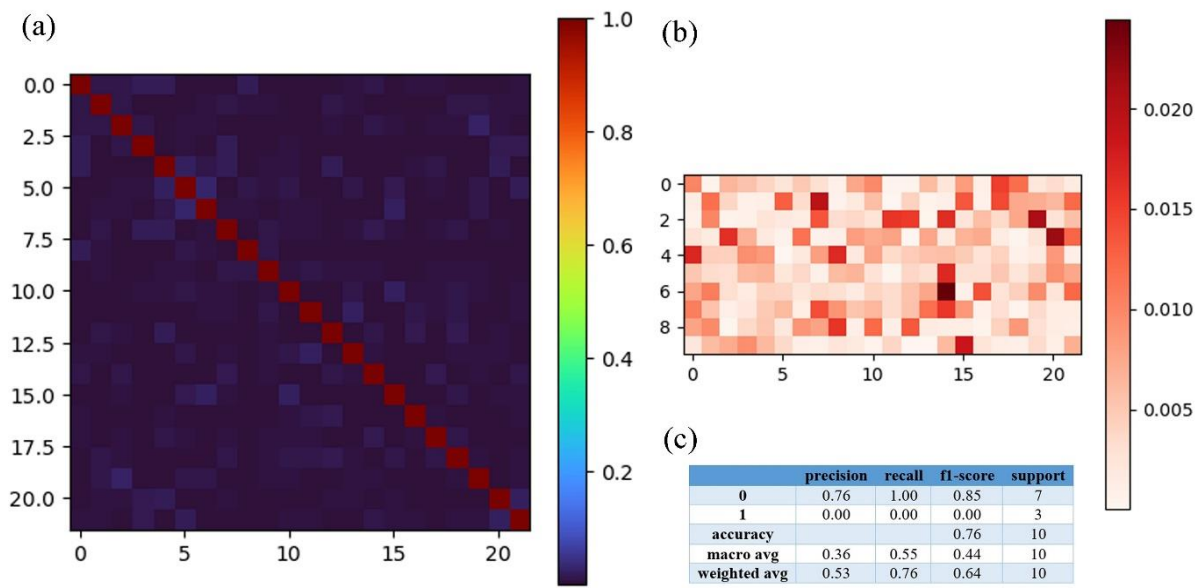

**Fig. 5.** Quantum kernel function on PCA reduced PSP-female dataset. **(a)** training kernel matrix, **(b)** testing kernel matrix, **(c)** classification report.

The aforementioned datasets are now executed on a real quantum computing backend (*IBM_osaka*) and on a simulator machine as described in the following section.

3.4 QML on the simulator and the real chip

The classification of each data point in a short-depth quantum circuit is error-mitigated while running on a real quantum computer [38] (Fig. 6). As the PCA-reduced PSP datasets contained eight important features, we have considered an encoding procedure in a four-qubit quantum computer in the IBM quantum experience platform [21]. Initially, we input a training set of data into the corresponding parameters and explored the probability distribution output by the quantum circuit (Fig. 6 a). We then allowed the circuit to operate on a quantum simulator (*qasm_simulator*) and together on a noisy simulator (*FakeManilaV2*) [39] that mimics a real machine to check how much the accuracy was to be improved. We run the model to generate quasi-probability distributions and measured the parameterized quantum circuit. The inherent nature of highly probabilistic computational results obtained needs us to have multiple runs ($shots = 1024$) for the same circuit to generate usable information in the form of probability distributions. The two primitives, which are (a) sampling probability distributions and (b) estimation of a value, are henceforth referred to as the sampler and the estimator, respectively defining the *QiskitRuntimeService* instance to run the program on the IBM quantum channel. Without error mitigation, the sampler returned a probability distribution corresponding to the measured samples (Fig 6 (b) (i)). The probability distribution was also obtained using a noisy simulator that uses a noise model with a fake backend class from the fake provider. The fake backends were developed to simulate the behaviors of IBM Quantum systems using system snapshots containing significant statistics that include qubit properties, basis gates, coupling map, error rate, and similar others that are relevant to achieve a noisy simulation of the quantum system. Accordingly, *FakeManilaV2* was imported to make the noise model. With this model, several data that might be configured such as noise model, basis gates, and coupling map, were set in the simulator option with a resilience level of 0. Further, to reduce the consequences of noise and decoherence, an error correction protocol was applied to negate the noise effects completely on an encoding level. Here we have used the matrix-free measurement mitigation (M3) routine to obtain better information quality [38, 39]. In the sampler, M3 was activated when we set the resilience level to 1 [39]. The details are depicted in Fig. 6 b and 6 c. The last plot (Fig. 6 c (i)) displays how error mitigation protocol can reduce the errors and bring on a probability distribution that is remarkably closer to the ideally simulated one. However, in the case of a real quantum computer, it is necessary to choose the least busy real backend (*IBM_osaka*) for the fastest execution. Thus, throughout the computation, we have used the IBM_quantum channel for all our run.

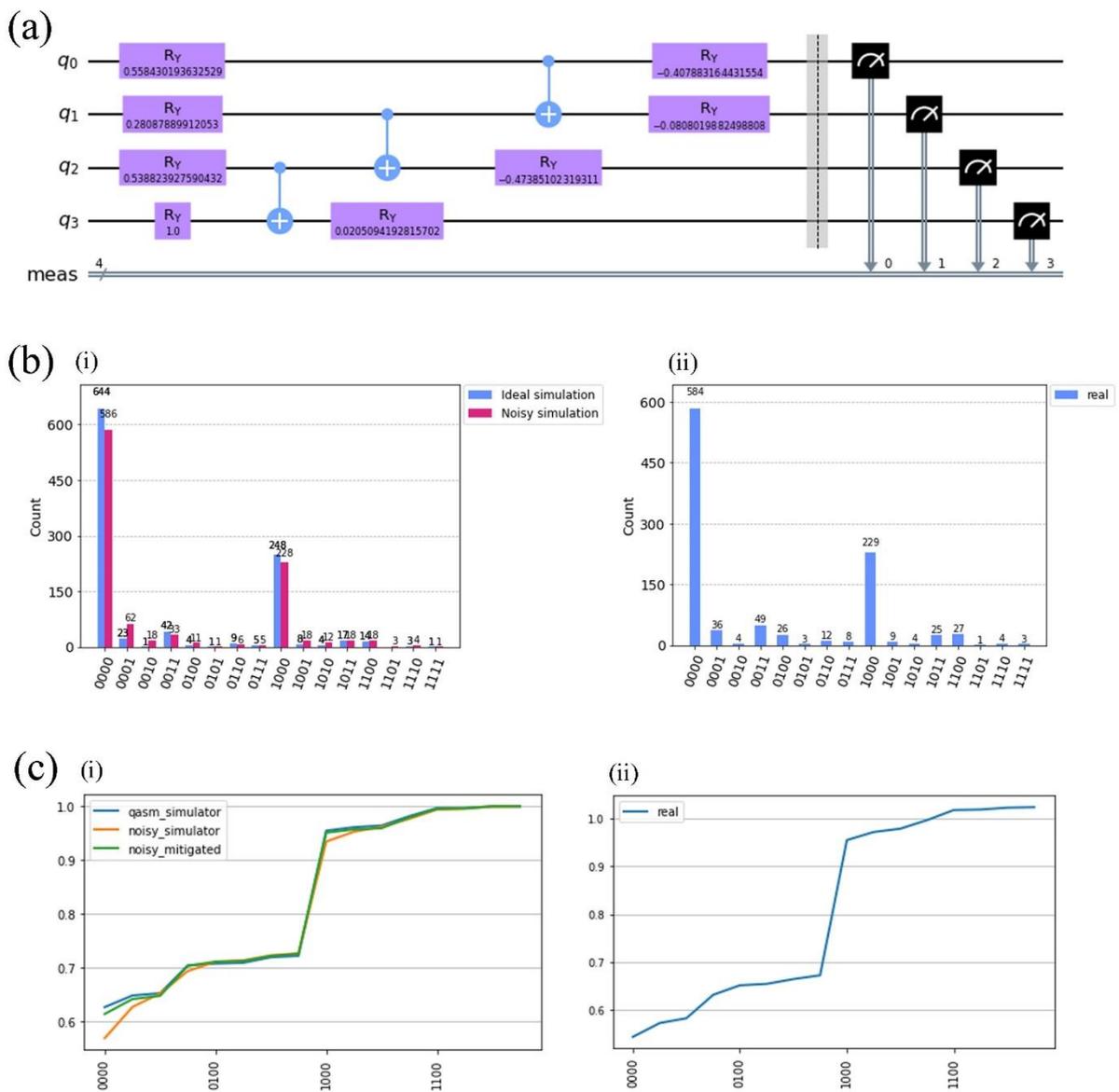

**Fig. 6.** Quantum machine learning on the simulator and real chip. **(a)** probability distribution output by the circuit on a training data, **(b)** probability distribution count using (i) qasm_simulator and noisy simulator and (ii) real backend computer, **(c)** cumulative probabilities of (i) qasm_simulator, noisy simulator and the simulator after error mitigated and (ii) real backend machine (*IBM_osaka*).

The above approach illustrates the ways to deal with quantum noise when applying a QML in the near-time quantum computing era. In the next section, fewer brain ROIs are identified that primarily correspond to the prefrontal, somatosensory association cortex, paracingulate, insula, deep cerebellar nuclei, etc., regions to differentiate PSP and HC groups.

### 3.5 Discriminative brain ROIs for diagnosis

The variations in PSP brain ROIs may be evaluated through the PSP rating scale (PCA-loadings), which considers symptom extremity for day-to-day behavior, activities, and functions, in brain regions like salience (insula, supramarginal left), limbic (mesothalamic junction, thalamus, anterior cingulate left), executive control (paracingulate, middle frontal gyrus), left basal ganglia, brainstem (rostral pons) and cerebellum (dentate nucleus right) networks respectively (Table 2). We next followed a hypothesis testing to understand whether the ROIs taken from two groups (i.e., PSP and HC) were consistent with a changed level of factors of interest. For this purpose, we have counted the number of times an $X_i$ from the PSP group was greater than a $Y_j$ from the HC group. The number was denoted by $U_{PSP}$. Similarly, the number of times an $X_i$ from PSP was smaller than a $Y_j$ from HC is denoted as $U_{HC}$. Therefore, the total number of pairwise comparisons that could be made is $N_X * N_Y$. Consequently, the hypotheses according to the *Mann-Whitney U* test may be given as: (i) the null hypothesis ($H_0$), when the two populations are equal, (ii) the alternative hypothesis ($H_1$) occurs when the two populations are not equal. Under the null hypothesis, we would expect $U_{PSP}$ and $U_{HC}$ to be approximately equal. This is majorly useful when the variance among two independent subject groups with a low number of counts in each subject (usually less than 30) was assessed, which were not again normally distributed. All of the PSP classifying nodes had significantly lower $p$−values when examined with a *Mann-Whitney U* test ($p < 0.05$) (Table 2). Therefore, we reject the hypothesis $H_0$ in favor of the alternative that the distributions were different. Alternatively, to classify PSP subjects compared to healthy control groups, pairwise ROIs were also examined. For example, brain ROIs were examined in the left middle frontal gyrus, right dentate nucleus, thalamus, and insula cortex that is compromised in PSP-female subjects. In HC, these regions contained rostral pons, left anterior cingulate, and paracingulate junction. The necessary details for all PSP subjects are shown in supplementary Fig. S4. Table 2 showed that PSP subjects include major classifying regions in the prefrontal, somatosensory association, paracingulate, insula cortex, brainstem, and within deep cerebellar nuclei and midbrain respectively, which is also depicted in Fig. 7. Based on functional MRI inputs and following preprocessing steps like EVC, PCA, and quantum as well as classical SVM classifiers, we aimed to assess patient subject groups compared to their healthy control ones with mild to moderate clinical disorder.

**Table 2**: Different predictive brain regions for PSP disease identification

| Nodes | Regions full names | loadings | p value $U = min(U_{PSP}, U_{HC})$ $U_X = N_X * N_Y + \frac{N_X(N_X+1)}{2} + \sum r_X$ | Network names |
|---|---|---|---|---|
| | | *female* | | |
| Lparcing | Left paracingulate | 0.2306 | 0.0011 | executive control |
| RIns | Right insula | 0.2055 | 0.018 | salience |
| LIns | Left insula | 0.1887 | 0.0068 | salience |
| rPons | Rostral pons | 0.1809 | 0.0483 | brainstem |
| RDentN | Right dentate nucleus | 0.1599 | 0.0363 | cerebellum |
| LACC | Left anterior cingulate | -0.1246 | 0.0082 | limbic |
| LMFG | Left middle frontal gyrus | 0.1173 | 0.029 | executive control |
| LThal | Left thalamus | 0.1162 | 0.0117 | limbic |
| RThal | Right thalamus | 0.1085 | 0.0075 | limbic |
| *male* | | | | |
| LMTJ | Left mesothalamic junction | -0.2201 | 0.0002 | limbic |
| LBG | Left basal ganglia | -0.2101 | 0.0154 | basal ganglia |
| RMTJ | Right mesothalamic junction | -0.2069 | 0.0006 | limbic |
| RMFG | Right middle frontal gyrus | -0.1978 | 0.027 | executive control |
| Rparacing | Right paracingulate | -0.1908 | 0.0408 | executive control |
| RIns | Right insula | -0.188 | 0.0022 | salience |
| LIns | Left insula | -0.1843 | 0.0057 | salience |
| RThal | Right Thalamus | -0.1776 | 0.0108 | limbic |
| RDentN | Right dentate nucleus | -0.1572 | 0.0016 | cerebellum |
| LSMG | Left supramarginal gyrus | 0.1536 | 0.0044 | salience |
| rPons | Rostral pons | -0.1364 | 0.0369 | brainstem |

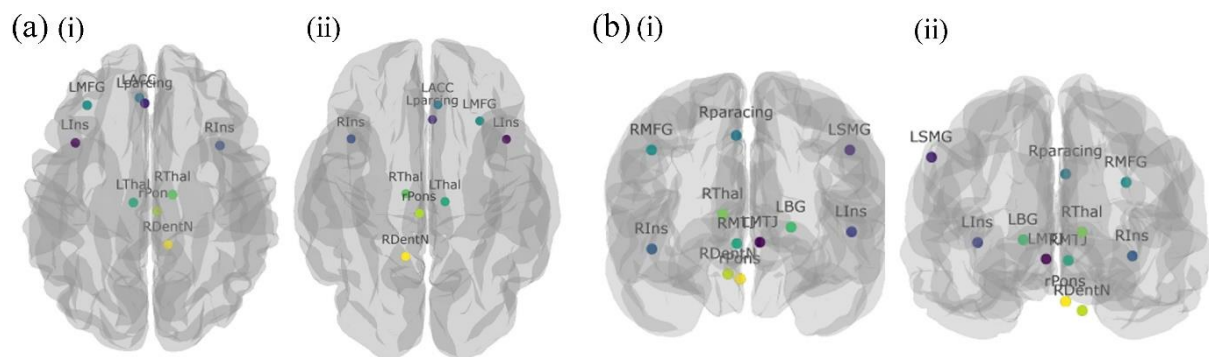

**Fig. 7.** Discriminative brain ROIs for PSP diagnosis. **(a)** PSP-female brain regions: (i) top view and (ii) bottom view, **(b)** PSP-male regions: (i) front view and (ii) back view.

## 4  Discussion

The present work describes and compares the performances of various quantum classifiers, which were formulated based on dimensionality reduction techniques of the underlying datasets. Using the standard definitions from the age-old graph theory, we have primarily attempted to transform the two-dimensional weighted adjacency matrix of the whole brain network to a one-dimensional EVC string. Subsequently, the EVC string was analysed according to Principal Component formalism, and eight principal components over twenty-seven brain ROIs were identified to ensure the maximum number of independent (i.e., orthogonal) states. Next, we have computationally explored the benefits of QVC (or ansatz) in initial ground-state preparations. We displayed predictable improvement as compared to various classical SVM algorithms and also to quantum kernels which were considered for comparison purposes. Without further showing the cases of trainability at random qubit sizes, we hypothesize that the present procedures will further promote the scope of such problems that can be addressed with near-term quantum devices.

The QVC designed in this study focuses on training data oracle that is parameterized by $\theta$, fed to the $W(\theta)$ block and further tuned and tested providing optimum $\theta$ measurements for the said classifier (i.e., *COBYLA* optimization). The algorithm attained an accuracy of 86%, analogous to that realized in the classical settings using an SVM and random forest classifier. The algorithm also signifies that the quantum ML procedure is capable of suitably classifying the pertinent information of the PSP dataset. For an ansatz depth of 2, much lesser computational resources were required to construct the quantum circuit. The merit of different QVC optimization was previously measured in the framework of QML classification tasks [11, 31]. Imperatively, in both QVC and kernel simulation of the present work, distinctive brain ROIs appeared and could certainly be accounted for in terms of the eight-qubit ansatz. Generally, larger dimensional datasets present substantial challenges for SVM and demand for computationally intensive tools as well. Therefore, quantum computers for ML with algorithms that are computationally intensive aiming to have outcomes in a reasonable time are likely to be needed. Moreover, it is also necessary to simulate QVC with *IBM-Qiskit* software. Accordingly, the environment, that is to be taken into consideration to run these tests, must be ideal. We have tested the circuit on a quantum simulator (*qasm_simulator*), a noisy simulator (*FakeManilaV2*), and a real backend machine (*IBM_osaka*). With the present day small-sized quantum machines and their comparatively elevated noise levels, it is difficult to perform large-scale computational problems and thus, the benefit of quantum computers remains unachievable. Nevertheless, an effective prototype model with improved accuracy concerning classical algorithms that a quantum classifier could attain, on a dataset must be low dimensional. Following the hypothesis, the QML's outputs were made superior to the respective classical ML algorithm by use of only an eight-component PCA-reduced dataset from the original adjacency matrix of 27 ROIs. Thus, the present QML exhibited improved effectiveness concerning the number of trainable constraints. As QML further develops, the accessibility of cloud-based quantum computers will increase for various classification tasks as demonstrated here.

So far, studies based on functional connectivity layouts of PSP have still been merely unpredictable [22, 40], though most studies have reported nearly consistent findings. We have identified 15 ROIs based on the respective scores to differentiate the target groups (i.e., PSP and HC) as shown in Table 2. Initially, using EVC, we calculated the average centrality measures of all input sets for each 27 brain ROIs. We next apply PCA to the resulting EVC series and use it as a covariance of interest in a statistical non-parametric measurement to obtain biomarkers for each subject. Brain ROIs with significant PCA loadings were compared between PSP and HC groups using *Mann-Whitney U* tests to identify regions with $p < 0.05$. Fig. 7 illustrates these 15 most differentiating brain ROIs. We found that 6 out of 15 ROIs are located in the prefrontal region, which signifies the clinical importance of the prefrontal region in PSP pathophysiology. A previous study found the signatures of PSP to be most prominent in the prefrontal-paralimbic region and brainstem [22, 40]. In addition to the prefrontal region, the present study also showed the presence of compromised ROIs within limbic and salience networks, while observable differences are also localized on the right supramarginal gyrus, insula, left paracingulate, and mesothalamic junction. Network studies on HC subjects have shown that the connections through basal ganglia primarily correspond to the subcortical-frontal lobe interface and any slight alteration can lead to executive and motor impairments, such as gait instabilities and falls [41]. Likewise, the dorsolateral prefrontal area of the executive control network, characterized by the middle frontal gyrus node signified the working memory region. Subsequently, the RMFG had impaired in PSP-male subject groups. Another two regions of the executive control network in the dorsal medial frontal cortex, namely, the ACC and paracingulate nodes are crucial for response selection as in Ref. [42]. In summary, we have noticed a scattered set of network ROIs that are also distributed in the brainstem, thalamus, insula, basal ganglia, deep cerebellar nuclei, and midbrain regions in addition to the prefrontal cortex are supposed to be affected in different scales for PSP subjects; our observation confirmed the outcomes of several neuropathological studies in PSP [40, 43]. The non-invasive and observable nature of network-based QML algorithms enabled us to identify potential signatures of PSP from the raw fMRI data. We apprehend future studies to assess such neurobiological signatures using QML models to explore and deliver conclusive evidence for automated classification of fMRI data.

## 5   Conclusion

In this study, we perform simulations on the public PSP dataset, which shows that QML on graph representations accomplishes promising performance output in fMRI data classification related to conventional methods, despite lacking several encounters, such as limited data and insufficient learning. The two classifiers – a variational quantum classifier and a quantum kernel estimator – proceed with the perception that a feature map that is difficult to evaluate classically is a fundamental point of forming a quantum benefit. In addition, we explore our present learning algorithms on state-of-the-art quantum computers made available by the IBM quantum experience platform in both simulator and real chip mode. We also discuss the interpretability of our preprocessing steps and find the discriminative brain ROIs and correlations for diagnosis. Given the ubiquity of QVC and other methods in machine learning, the present technique may encompass approaches beyond binary classification and also highlight that error-mitigation methods suggest a way to precise classification of different brain disorders even with NISQ hardware.

**Data availability statement**:

The data that support the findings of this study are openly available in USC Multimodal Connectivity Database (UMCD) (http://umcd.humanconnectomeproject.org/).

**Conflict of interest statement**

The author declares no conflicts of interest. This research did not receive any specific grant from funding agencies in the public, commercial, or not-for-profit sectors.

**Declaration of Generative AI and AI-assisted technologies in the writing process**

The author did not use any AI tools to analyse and draw insights from data as part of the research process.

Supplementary Section of Quantum Machine Learning with Application to Progressive Supranuclear Palsy Network Classification

**Table S1.** Details of the nodes *ROI* (i.e., region name, node location in terms of its functional network, and MNI coordinates)

| Nodes | | Network | MNI coordinates | | |
|---|---|---|---|---|---|
| **Region abbreviation** | **Region full names** | | x | y | z |
| LMFG | Left middle frontal gyrus | executive control | -32.4058 | 30.89855 | 32.11594 |
| Lparcing | Left paracingulate | executive control | -2.9 | 31.8 | 32.3 |
| LSMG | Left supramarginal gyrus | salience | -54.252 | -46.3465 | 41.00787 |
| LPreCu | Left precuneus | default mode | -4.46296 | -73.0741 | 39.53704 |
| LRSC | Left retrosplenial cortex | limbic | -7.01587 | -52.6349 | 9.650794 |
| LACC | Left anterior cingulate | limbic | -5.91667 | 37.91667 | 17.20833 |
| LpMCC | Left posterior midcingulate | limbic | -2.31579 | -28.6316 | 38.94737 |
| LIns | Left insula | salience | -43.4 | 15.25 | 0.85 |
| RMFG | Right middle frontal gyrus | executive control | 39.22124 | 26.1062 | 34.63717 |
| Rparacing | Right paracingulate | executive control | 4.983051 | 32.20339 | 39.83051 |
| RSMG | Right supramarginal gyrus | salience | 60.37895 | -47.7053 | 27.2 |
| RPreCu | Right precuneus | default mode | 10.33766 | -69.2208 | 38.8052 |
| RRSC | Right retrosplenial cortex | limbic | 6 | -54.4381 | 18.59048 |
| RACC | Right anterior cingulate | limbic | 10.2029 | 35.13044 | 19.50725 |
| RpMCC | Right posterior midcingulate | limbic | 2.059406 | -28.7723 | 38.13861 |
| RIns | Right insula | salience | 41.16279 | 14.51163 | -6.51163 |
| LBG | Left basal ganglia | basal ganglia | -18.1767 | 10.90763 | 2.586345 |
| LThal | Left thalamus | limbic | -9.71795 | -19.3846 | 6.487179 |
| RBG | Right basal ganglia | basal ganglia | 17.6063 | 10.17323 | 3.944882 |
| RThal | Right thalamus | limbic | 12.42953 | -14.8859 | 7.422819 |
| LMTJ | Left mesothalamic junction | limbic | -5.2973 | -13.4054 | -5.94595 |
| RMTJ | Right mesothalamic junction | limbic | 5.64486 | -13.7009 | -6.61682 |
| rPons | Rostral pons | brainstem | 4.184615 | -24.9538 | -25.2 |
| cPons | Caudal pons | brainstem | 1.632653 | -32.8571 | -36.2449 |
| LDentN | Left dentate nucleus | cerebellum | -11.5556 | -49.8889 | -24.9444 |
| RDentN | Right dentate nucleus | cerebellum | 11.04615 | -46.3692 | -25.7538 |
| Vermis | Vermis | cerebellum | 0.926316 | -51.6 | -14.2316 |

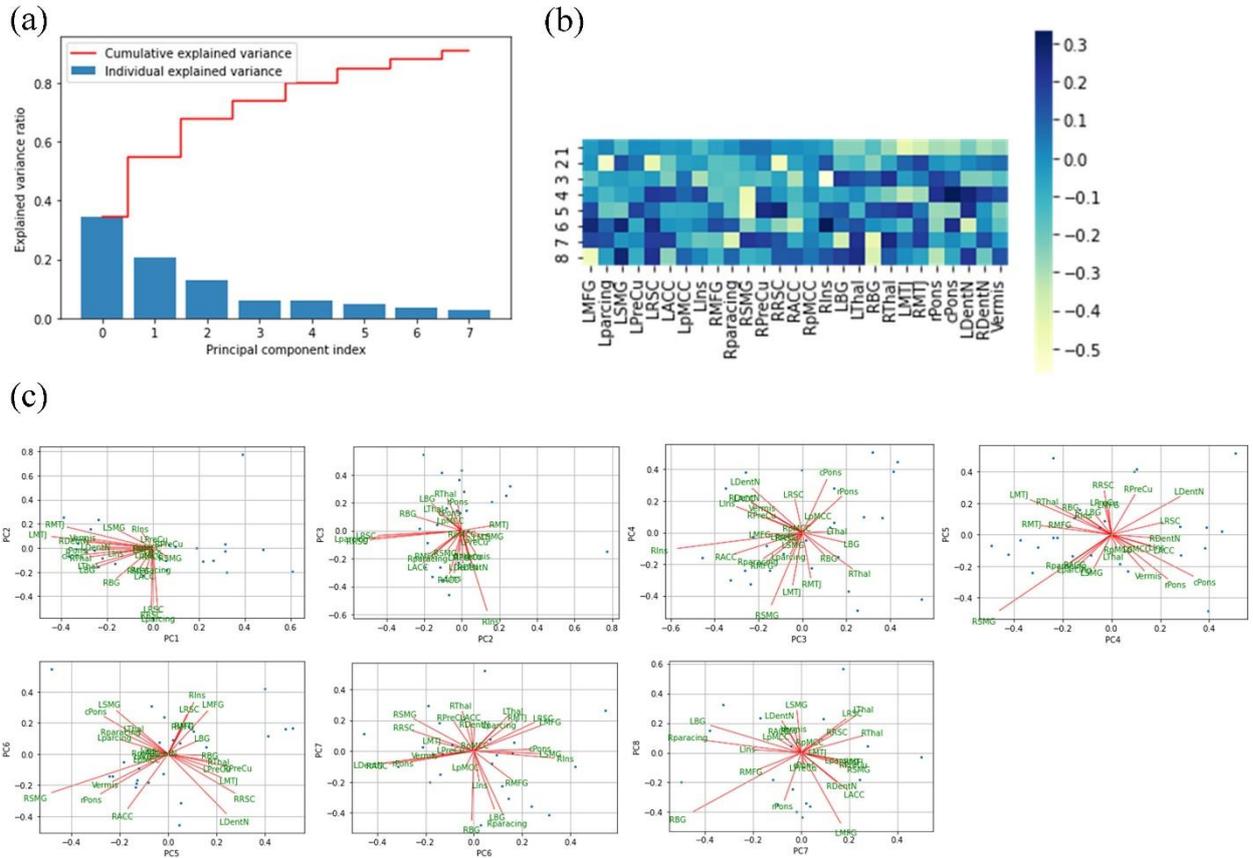

**Fig. S1.** PCA on EVC reduced PSP male dataset. **(a)** Scree plot for proportion of variance explained, **(b)** Correlation heatmap among 8 principal components distributed over 27 brain ROIs, **(c)** Variances of ROIs on the factorial plane. Each plane consisting of the intersection of two of the principal components.

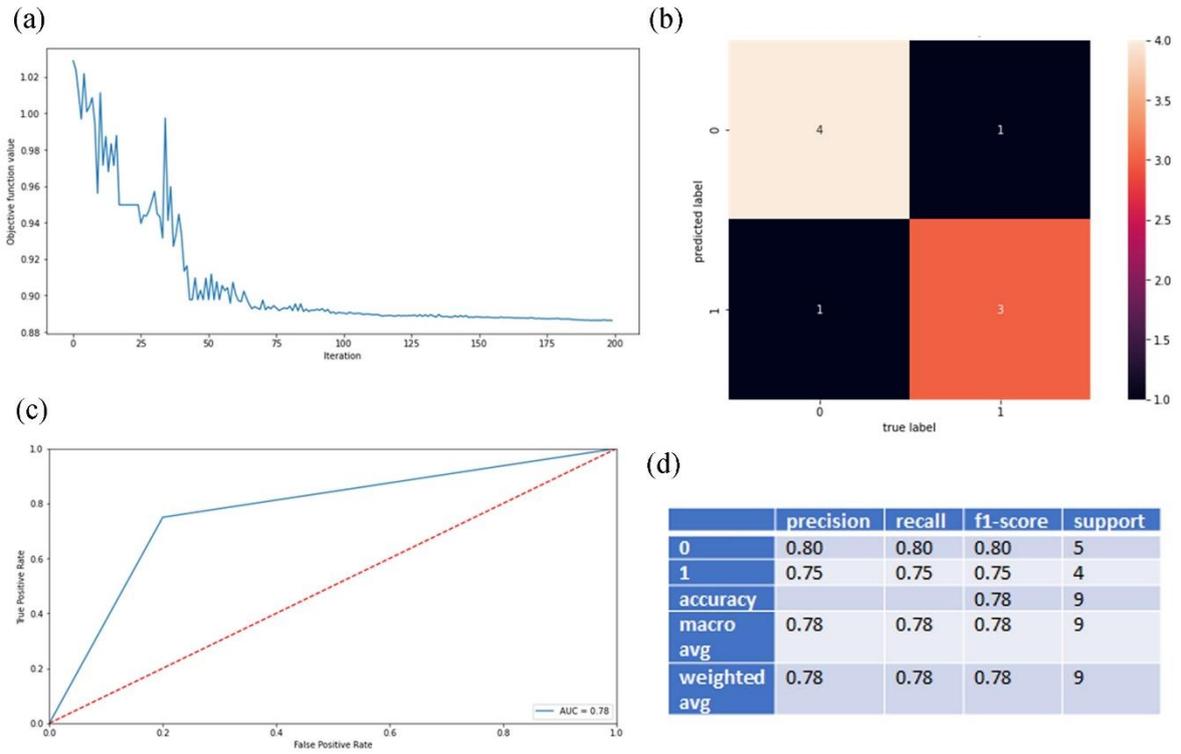

**Fig. S2.** QVC after PCA reduced PSP male dataset. **(a)** COBYLA optimizer with $R = 200$ shots, **(b)** Confusion matrix of QVC, **(c)** ROC graph for variational quantum circuit, **(d)** classification report.

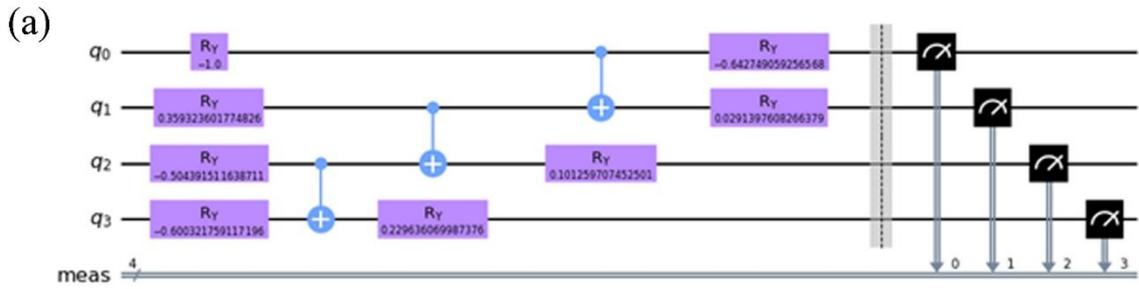

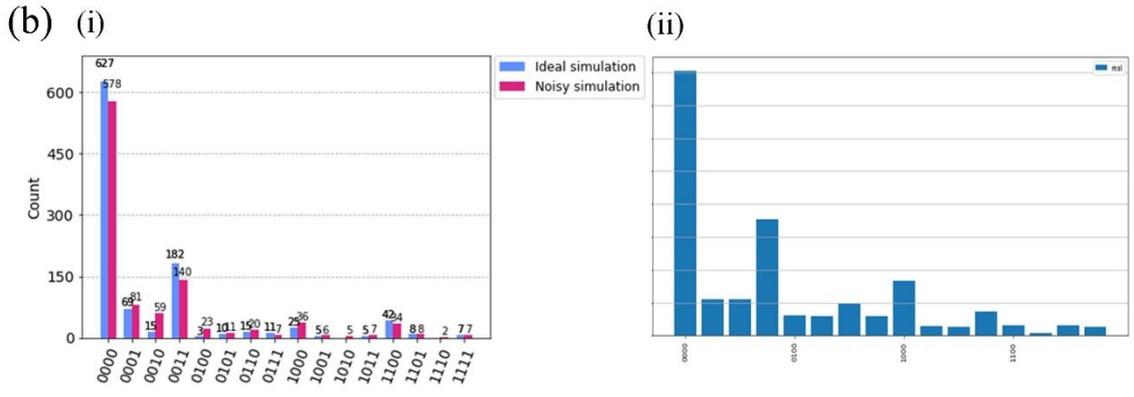

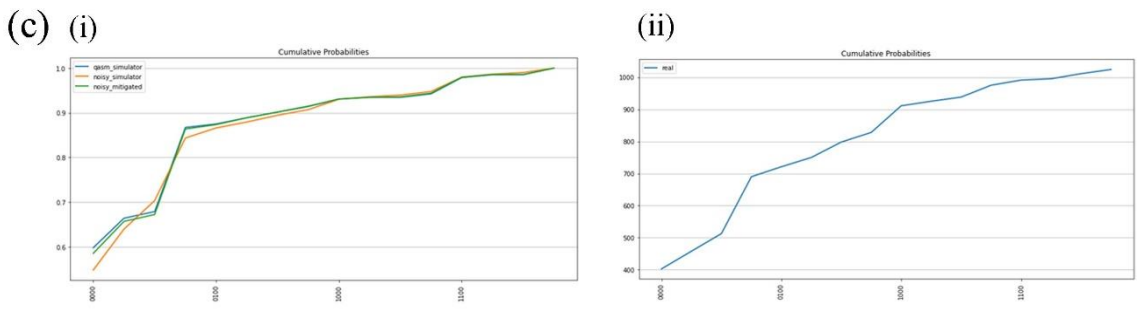

**Fig. S3.** Quantum machine learning on the simulator and real chip of PSP male dataset. **(a)** probability distribution output by the circuit on a training data, **(b)** probability distribution count using (i) qasm_simulator, noisy simulator and the simulator after error mitigated and (ii) real backend computer, **(c)** cumulative probabilities of (i) qasm_simulator, noisy simulator and the simulator after error mitigated and (ii) real backend machine (ibm_kyoto).

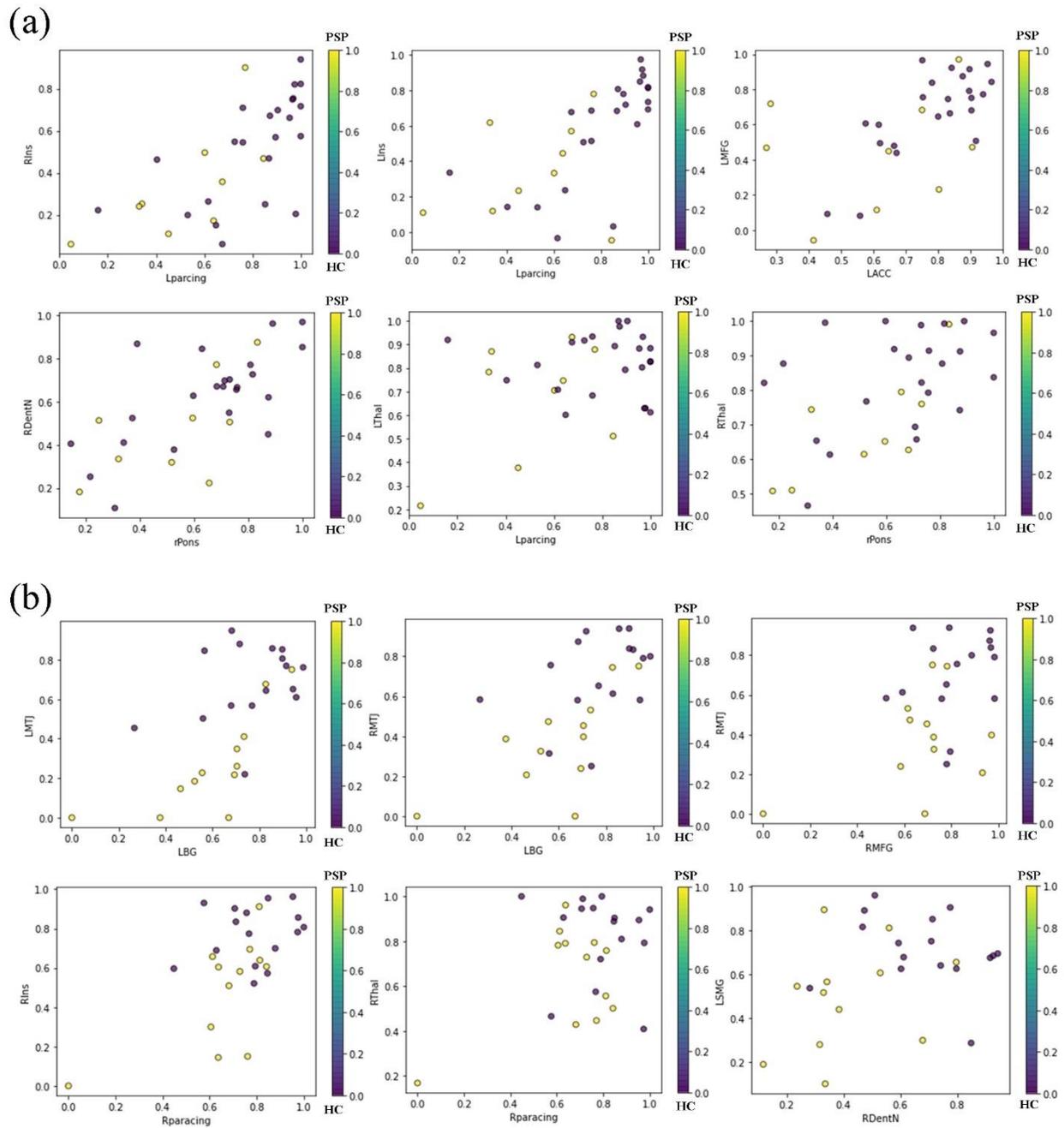

**Fig. S4.** Pairwise comparison of selective ROIs for classification between PSP and HC. **(a)** PSP-female subjects, **(b)** PSP-male subjects.